# First-principle calculations on Li$_2$CuSb: A novel material for lithium-ion batteries

A. Shukla[1], S. Pandey[2], and H. Pandey[2,*]
[1]*Department of Physics, Integral University, Lucknow, Uttar Pradesh, India.*
[2]*Department of Physics, Sardar Vallabhbhai National Institute of Technology, Surat-395007, India.*[*]

We investigate the Li$_2$CuSb full-Heusler alloy using the first-principles electronic structure calculations and propose the electrochemical lithiation in this alloy. Band structure calculations suggest the presence of metallic nature in this alloy contrary to half-metallic nature as predicted for most of the members of the full-Heusler alloy family. This alloy is found to be a promising anode material for high-capacity rechargeable batteries based on lithium-ion. We found a removal voltage of $\approx$ 2.48 V for lithium ions in the Li$_2$CuSb/Cu cell, which is in good agreement with the experimentally obtained result for a similar kind of material Cu$_3$Sb. During charge and discharge cycles of the Li$_2$CuSb/Cu cell, the formation of a non-stoichometric compound (Li$_{2-y}$Cu$_{1+x}$Sb) having a similar structure as Li$_2$CuSb suggests a better performance as well as stability of this cell.

## I. INTRODUCTION

Rapidly growing demand, ever-increasing power costs, limited natural sources, environmental pollution, global warming, *etc* are some of the global issues, on which almost every country is pumping in a huge amount of funding for research and development. Apart from energy generation, its storage and harnessing is also an equally important issue, and to fulfill the quest of a modern society filled with numerous kinds of gadgets and devices, a variety of novel materials are being explored for various applications such as rechargeable batteries. Various researchers around the globe are working very hard to design and develop low-cost new materials for compact, flexible rechargeable batteries. For the sake of better air quality in the cities and to lower the dependency on fossil fuels, the demand for electric vehicles is continuously increasing. The development of new materials will definitely open the possibility of more durability and higher capacity of batteries. Many graphite-based materials have been extensively investigated and used as anode for lithium-based batteries, but the irreversible capacity loss due to electrolyte decomposition and structural change prohibits their applications [1]. Several other types of compounds and composites based on tin [2–5] and silicon have been extensively studied [6–9]. The utilization of nanomaterials such as nanoparticles and nanowires to minimize the volume expansion have also been tried [10–12]. During the lithiation process, the usage of transition metal oxides as an anode has also exhibited a better cycling stability [13, 14]. Another versatile material, known as Heusler alloys, is mostly investigated for spintronics [15–18] thermoelectrics [19–23], catalysis [24–26], and many more interesting areas [27–29], very few people have tried to explore these alloys as a potential electrode material for lithium ion-based batteries [30–32]. The better stability, robustness, sustainability, and a high value of theoretical gravimetric (specific) capacity make these materials a promising candidate for battery applications.

Heusler alloys are classified in mainly two categories: Half-Heusler alloys (XYZ, 1:1:1) and full-Heusler alloys (X$_2$YZ, 2:1:1), where the atoms X and Y are the transition metals, while Z is either a semiconductor or a non-magnetic metal and so far more than 1600 members in Heusler alloys family have been investigated. Apart from these, inverse Heusler alloys and quarternary Heusler alloys (XX'YZ) are also well explored. The unit cell consists of 4 interpenetrating *fcc* sublattices, where the Wyckoff positions of the atoms are given by X: 4*a* (0,0,0) & 4*b* (0.5,0.5,0.5); Y: 4*c* (0.25,0.25,0.25) and Z: 4*d* (0.75,0.75,0.75). The absence of one atom from the X-site will create semi- or half-Heusler alloys.

Co$_2$MnSi Heusler alloy is one of the most investigated materials in the field of the spintronics industry and the same has also been investigated for lithium batteries as an anode material [30, 31]. In another alloy from this family, a discharge capacity of 220 mAhg$^{-1}$ during the first charge cycle has been demonstrated through the formation of lithiated Li$_2$CoSb phase in the Heusler alloys CoMnSb [32]. Modern techniques such as machine learning are also being carried out in Heusler alloys to predict new materials with desired properties [33, 34]. On the other hand, very few theoretical investigations have been reported on Heusler alloys in order to make use of them as an electrode for high capacity rechargeable batteries [35]. In this work, we report first-principle calculations on Li-based full-Heusler alloy Li$_2$CuSb (LCS) and investigate the bandstructure as well as density of states (DOSs) for the optimized unit cell, and further propose the lithiation/delithiation processes in LCS/Cu electrochemical cell. A systematic study of the electronic properties of these alloys has already proven useful for various applications.

## II. COMPUTATIONAL DETAIL

First principle calculations were performed on full-Heusler alloy LCS using two different codes for better structure optimization. We have used Green's function formalism-based full potential SPRKKR method for the structural optimization [36]. Various self-consistent calculations for different potentials have been carried out for a $k$-grid of 22× 22× 22 and 30 energy points on the complex energy path. An initial value of lattice parameter of 6.28 Å was used. These calculations were carried out on LCS with fcc structure (space group: 225, $Fm\bar{3}m$) where Li atoms are at (0.25,0.25,0.25) and (0.75, 0.75, 0.75); Cu atoms at (0, 0, 0), and Sb atoms at (0.50,0.50,0.50) atomic positions. For self-consistent calculations (SCF), the energy convergence criterion was set to be at $10^{-5}$ Ry.

---

[*] Author to whom correspondence should be addressed. Electronic mail:hp@phy.svnit.ac.in

## III. RESULTS AND DISCUSSION

Firstly, SCF calculations have been performed on full-Heusler alloy (LCS) with a lattice parameter of 6.28 Å. After attaining the convergence criterion, volume optimization is done to obtain an optimized lattice parameter. For these calculations, the isotropic strain has been applied which maintains the unit cell geometry and the formula unit cell volume has been varied in the range of ± 10%. To estimate the actual volume per formula unit, total unit cell volume has been divided by 4. Figure 1 shows the variation of total energy (in Ry) with unit cell volume. The Murnaghan equation of state is used to obtain energy minimum and a minimum in unit cell volume is found to be at 432.47 au$^3$ (au = atomic unit) and the corresponding equilibrium lattice parameter is 6.352 Å. After this, all further calculations have been done using this equilibrium lattice parameter.

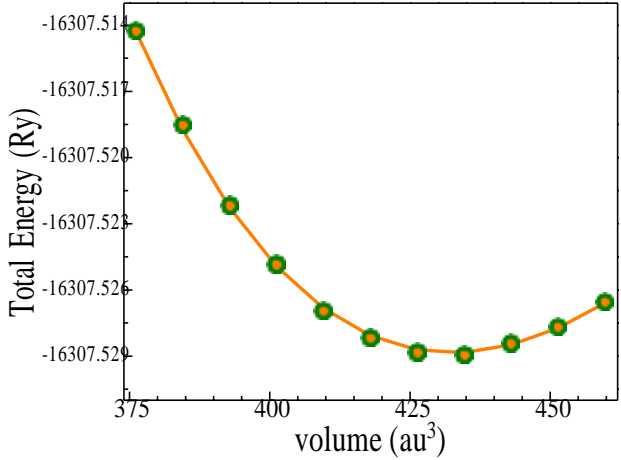

FIG. 1. (Color online) The total energy as a function of unit cell volume of Li$_2$CuSb along with its fitting using the Murnaghan equation of state.

Firstly, to get the information about chemical bonding between different atoms, we have calculated the atom orbital projected local density of states (PDOS) and total density of states (DOS) for LCS. Figures 2-5 show the total DOS as calculated for $L2_1$ ordered LCS Heusler alloy, PDOS for Cu, Sb, and Li atoms, respectively. The total DOS is dominated by the PDOS of Cu-$d$ and Sb-$s$ states, wherever $p$ states of the Li have very minimal contribution. Hence, the $d$ orbitals of Cu and Sb atoms are strongly hybridized as compared to other atomic orbitals. To have a better idea about DOSs and PDOSs, the Y-axis has been zoomed in. These DOSs are extended from -4.0 eV to $\varepsilon_F$ and corresponding contributions are 37.4 electrons/eV for Cu-$d$, 0.18 electrons/eV for Cu-$p$ states, 0.16 electrons/eV for Cu-$s$, 0.15 electrons/eV for Sb-$d$, 0.94 electrons/eV for Sb-$p$, 2.55 electrons/eV for Sb-$s$, 0.14 electrons/eV for Li-$p$, and 0.04 electrons/eV for Li-$s$. This observation suggests that some of the electrons are transferred to the valence band and contribute to a weak covalent interaction between the same type of atoms.

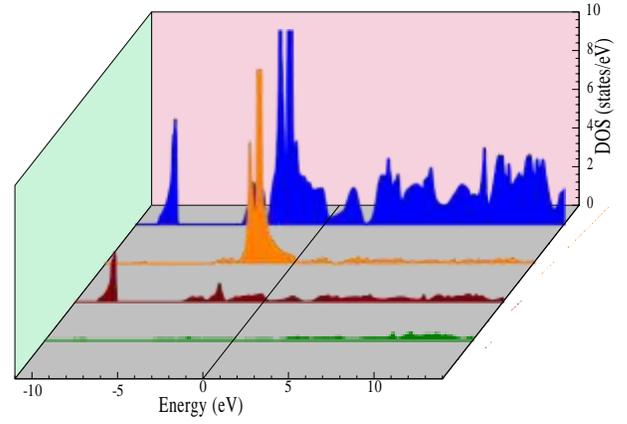

FIG. 2. (Color online) The total density of states of $L2_1$ ordered Li$_2$CuSb Heusler alloy along with partial DOSs for Cu, Sb, and Li atoms.

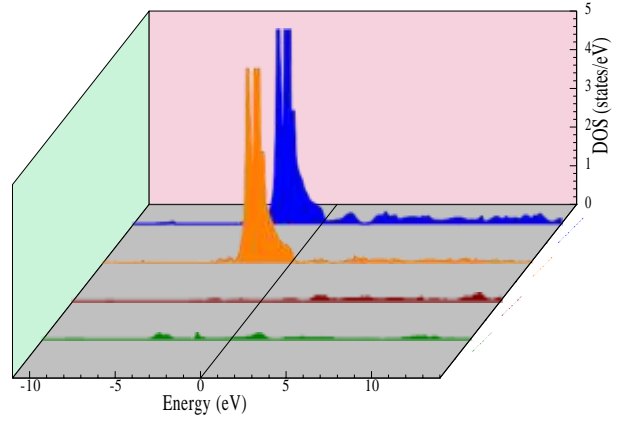

FIG. 3. (Color online) The total density of states of Cu atoms in Li$_2$CuSb Heusler alloy along with orbital projected DOSs.

Figure 6 shows the band structure plot for LCS Heusler alloy. A clear cross-over of a band (marked with black circles) suggests the metallic nature. The major occurrence of these bands near the Fermi energy ($\varepsilon_F$) results in the increased bandwidth which further pushes the conduction and valence bands towards lower and higher energies, respectively with respect to $\varepsilon_F$. It is also found that most of the hybridization between Sb-$s$ and Cu-$s/p$ states is occurring around energy ranging from -10.0 to -5.0 eV. On the other hand the $s$, $p$, and $d$ orbitals of Cu hybridize with $p$ and $d$ orbitals of Sb around -5.0 eV energy range. The bands corresponding to Li-$s$ and Li-$p$ states are found to be very broad. All of these strong electronic correlations suggest the strong hybridization in this lithiated Heusler alloy.

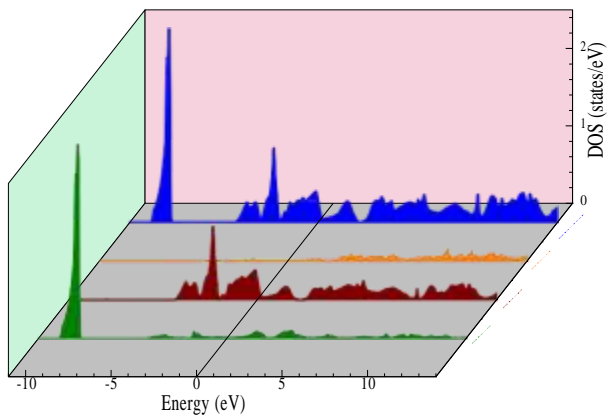

FIG. 4. (Color online) The total density of states of Sb atoms in Li$_2$CuSb Heusler alloy along with orbital projected DOSs.

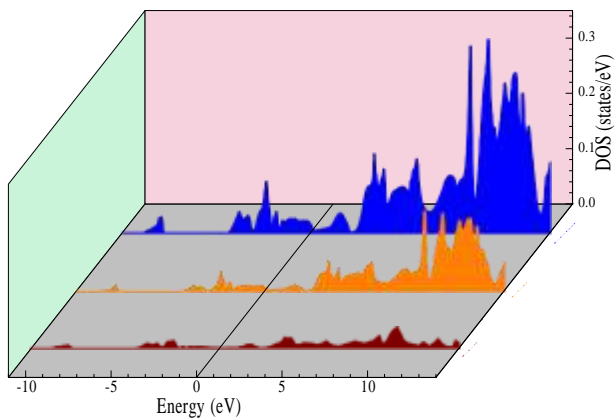

FIG. 5. (Color online) The total density of states of Li atoms in Li$_2$CuSb Heusler alloy along with orbital projected DOSs.

In order to understand the mechanism for the lithiation/delithiation processes and electrochemical properties of this proposed LCS heusler alloy, one of the possible interaction reactions can be written as:

$$\text{Li}_2\text{CuSb} + \text{Cu} \rightarrow \text{Cu}_2\text{Sb} + 2\text{Li}^+ + 2e^- \quad (1)$$

Here, from the above reaction, we can see that there are 2 electrons that are involved in the LCS/Cu electrochemical cell process and a homogeneous insertion and extrusion of Cu and Li, respectively can occur. Other possibility could be the formation of Li$_3$Sb with a transfer of single electron (for LCS/Li cell). In general, above reaction can be written as:

$$\text{Li}_2\text{CuSb} + x\text{Cu} \rightarrow \text{Li}_{2-y}\text{Cu}_{1+x}\text{Sb} + y\text{Li} \quad (2)$$

with $0 < x, y \leq 1$. From the total energy differences between reactants and products, we can get information about the average lithium removal voltage for the former case.

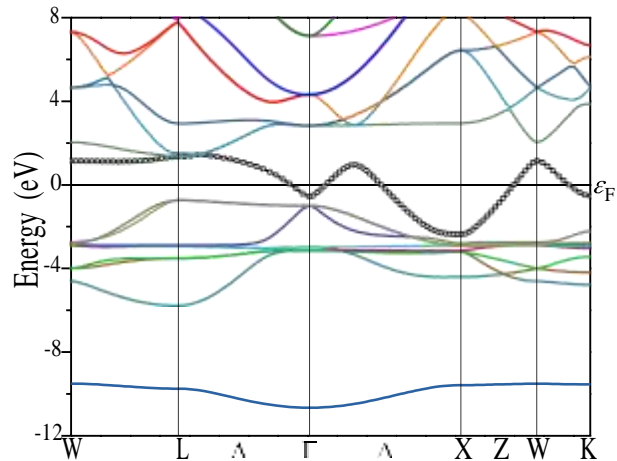

FIG. 6. (Color online) Calculated electronic band structure of Li$_2$CuSb Heusler alloy. The band which is marked with black circles crosses the Fermi level.

$$V = \frac{1}{zF}\left[\begin{array}{c} E_{total}(\text{Li}_2\text{CuSb}) + E_{total}(\text{Cu}) \\ -E_{total}(\text{Cu}_2\text{Sb}) - 2E_{total}(\text{Li}^+) \end{array}\right]$$

Here, E$_{total}$ represents the total energy per formula unit, $z$ is the electronic charge that is transported by lithium in the electrode, and $F$ is the Faraday constant ($\approx$96485$A$ $sec./mol$). By substituting all the required energies, we get a voltage of $\approx$ 2.48 V for the removal of lithium in Li$_2$CuSb/Cu, which is pretty close to the previously experimentally obtained value of 3.0 V for a similar kind of material Cu$_2$Sb [37]. During the discharge cycle, the lattice volume can contract without any abrupt changes and we propose the following chemical reactions via which the insertion and removal of Li atoms in this electrochemical cell can occur by a solid-state solution:

$$\text{Li}_2\text{CuSb} \rightarrow x\text{Li} + \text{Li}_{2-x}\text{CuSb} \quad (3a)$$
$$\text{Li}_{2-x}\text{CuSb} + x\text{Li} \rightarrow \text{Li}_2\text{CuSb} \quad (3b)$$

Here, the first reaction (3a) is for 1$^{st}$ discharge while the second one (3b) is for 2$^{nd}$ charge. Both the charge and discharge cycle are accompanied by the formation of Lithium-Copper-Antimony alloy in the same crystal structure as for LCS. The stability of this intermediate The theoretical specific gravimetric capacity for LCS is $\approx$ 269 mAh/g, so we can expect a high delivered rechargeable capacity of around 90% by considering a good utilization of this robust and dense alloy as an electrode. Based on the above proposal, we can expect some experimental work on this alloy could come up which will definitely trigger Heusler alloys to be used in this direction.

imental results of a similar kind of $Cu_2Sb$ cell. Similar structural relationships with LCS and Lithium-Copper-Antimony (formed during charging and discharging cycle reactions) alloys, provide better stability for LCS/Cu cells. Hence, this alloy may be a possible candidate for its applications in the field of lithium-based batteries, but more investigation in the experimental direction is required.

## IV. CONCLUSIONS

In this work, full Heusler alloy $Li_2CuSb$ has been investigated using density functional theory and by applying the Murnaghan equation, unit cell volume is optimized. We have predicted that the $Li_2CuSb$ full-Heusler should be a promising candidate for the electrode material to be used in high capacity rechargeable lithium-ion batteries. The prediction of lithium removal voltage of $\approx$ 2.48 V in $Li_2CuSb$/Cu cell, is in good agreement with the exper-

## V. DECLARATION OF COMPETING INTEREST

The authors declare that they have no known competing financial interests or personal relationships that could have appeared to influence the work reported in this paper.

## VI. ACKNOWLEDGEMENTS

This work is supported by the research grant from the Science and Engineering Research Board (SERB), Govt. of India, against scheme ECR/2017/001612.

[1] M. Inaba, Z. Siroma, A. Funabiki, and Z. Ogumi, Langmuir **12** 1534 (1996).
[2] Y. Idota, T. Kubota, A. Matsufuji, Y. Maekawa, and T. Miyasaka, Science **276** 1395 (1997).
[3] J. Zhang and Y. Xia, J. Electrochem. Soc. **153** A1466 (2006).
**[4]** F. Wang, M. Zhao, and X. Song, J. Power Sources **175** 558 (2008).
[5] J. Yin, M. Wada, S. Yoshida, K. Ishihara, S. Tanase, and T. Sakai, J. Electrochem. Soc. **150** A1129 (2003).
[6] I.-S. Kim, G.E. Blomgren, and P.N. Kumta, J. Power Sources **130** 275 (2004).
[7] G. X. Wang, L. Sun, D. H. Bradhurst, S. Zhong, S. X. Dou, and H. K. Liu, J. Power Sources **88** 278 (2000).
[8] H. Kim, J. Choi, H.-J. Sohn, and T. Kang, J. Electrochem. Soc. **146** 4401 (1999).
[9] H. Dong, X. P. Ai, and H. X. Yang, Electrochem. Commun. **5** 952 (2003).
[10] C. K. Chan, H. Peng, G. Liu, K. Mcilwrath, X. F. Zhang, R. A. Huggins, and Y. Cui, Nature Nanotechnol. **3** 31 (2008).
[11] D. Son, E. Kim, T. Kim, M. Kim, J. Cho, and B. Park, Appl. Phys. Lett. **85** 5875 (2004).
[12] G. Armstrong, A. R. Armstrong, P. G. Bruce, P. Reale, and B. Scrosati, Adv. Mater. **18** 2597 (2006).
[13] S. W. Oh, H.J . Bang, Y. C. Bae, and Y. K. Sun, J. Power Sources **173** 502 (2007).
[14] P. Poizot, S. Laruelle, S. Grugeon, L. Dupont, and J.-M. Tarascon, Nature **407** 496 (2000).
[15] H. Pandey, P. C. Joshi, R. P. Pant, R. Prasad, S. Auluck, and R. C. Budhani, J. Appl. Phys. **111** 023912 (2012).
[16] H. Pandey and R. C. Budhani, J. Appl. Phys. **113** 203918 (2013).
[17] P. K. Rout, H. Pandey, L. Wu, Anupam, P. C. Joshi, Z. Hossain, Y. Zhu, and R. C. Budhani, Phys. Rev. B **89** 020401(R) (2014).
[18] H. Pandey, P. K. Rout, Anupam, P. C. Joshi, Z. Hossain, and R. C. Budhani, Appl. Phys. Lett. **104** 022402 (2014).
[19] K. Xia, C. Hu, C. Fu, X. Zhao, and T. Zhua, Appl. Phys. Lett. **118** 140503 (2021).
[20] L. Huang, Q. Zhang, B. Yuan, X. Lai, X. Yan, and Z. Ren, Mat. Res. Bull. **76** 107 (2016).
[21] S. Singh and H. Pandey, Materials Today: Proceedings **28** 325 (2020).
[22] H. Pandey, S. Chhoker, B. C. Joshi, and D. Tripathi, AIP Conf. Proc. **2136** 040004 (2019).
[23] D. R. Jaishi, N. Sharma, B. Karki, B. P. Belbase, R. P. Adhikari, and M. P. Ghimire, AIP Advances **11** 025304 (2021).
[24] H. Liang, F. Chen, M. Zhang, S. Jing, B. Shen, S. Yin, and P. Tsiakaras, Appl. Cat. A, Gen. **574** 114 (2019).
[25] T. Kojima, S. Kameoka, S. Fujii, S. Ueda, and A.-P. Tsai, Sci. Adv. **4** eaat6063 (2018).
**[26]** T. Kojima, S. Kameoka, and A.-P. Tsai, ACS Omega **2** 147 (2017).
[27] V. Toutam, H. Pandey, Sandeep Singh, and R. C. Budhani, AIP Advances **3** 022124 (2013).
[28] H. Pandey, M. Kumar, A. K. Srivastava, and S. Pandey, AIP Conf. Proc. **2009** 020029 (2018).
[29] H. Pandey, R. Prasad, and R. C. Budhani, JPS Conf. Proc. **3** 017037 (2014).
[30] S. Matsuno, M. Noji, T.Kashiwagi, M. Nakayama, and M. Wakihara, J. Phys. Chem. C 111 7548 (2007).
[31] J. H. Park, D. H. Jeong, S. M. Cha, Y.-K. Sun, and C. S. Yoon, J. Power Sources **188** 281 (2009).

[32] S. Matsuno, M. NaKayama, and M. Wakihara, J. Electrochem. Soc. **155** A61 (2008).
[33] X. Hu, Y. Zhang, S. Fan, X. Li, Z. Zhao, C. He, Y. Zhao, Y. Liu, and W. Xie, J. Phys. Condens. Matter, **32** 205901 (2020).
[34] Y. Zhang and X. Xu, AIP Advances 10, 045121 (2020).
[35] A. H. Reshak and H. Kamarudin, J. Alloys and Compds. 509 **7861** (2011).
[36] H. Ebert, D. Ködderitzsch, and J. Minar, Rep. Prog. Phys. **74**, 096501 (2011).
[37] L. M. L. Fransson, J. T. Vaughey, R. Benedek, K. Edstrom, J. O. Thomas, and M. M. Thackeray, Electrochem. Commun. **3** 317 (2001).